\documentclass[epj,a4paper]{svjour}
\usepackage{multirow,rotating}
\usepackage[latin1]{inputenc}  %Ümläute
\usepackage{graphicx}% Include Figure files
\usepackage{siunitx}

\graphicspath{{../Figures-EPS/}{.}{}}
\usepackage[]{color}
\usepackage[]{graphicx}
\usepackage[]{enumerate}
\usepackage[]{lineno}
\usepackage[square,comma,numbers,sort&compress]{natbib} 

\newcommand{\infnMi}{4}
\newcommand{\infnpd}{1}
\newcommand{\infnLNL}{2}
\newcommand{\DipPD}{3}
\newcommand{\turkey}{5}

\newcommand{\Al}{$^{26}$Al }
\newcommand{\MgAl}{$^{25}$Mg($\alpha$,n)$^{28}$Si }

\begin{document}

\title{A new study of \MgAl angular distributions  at $E_\alpha$ = 3 - 5 MeV}
\author{
	A.\,Caciolli \inst{\infnpd , \infnLNL}\thanks{\emph{corresponding author:} caciolli@pd.infn.it} \and
	T.\,Marchi \inst{\infnLNL , \DipPD} \and
	R.\,Depalo \inst{\infnpd, \DipPD} \and
	S.\,Appannababu \inst{\infnLNL} \and
	N.\,Blasi \inst{\infnMi} \and
	C.\,Broggini \inst{\infnpd} \and
	M.\,Cinausero \inst{\infnLNL} \and
	G.\,Collazuol \inst{\infnpd , \DipPD} \and
	M.\,Degerlier \inst{\turkey} \and
	D.\,Fabris \inst{\infnpd} \and
	F.\,Gramegna \inst{\infnLNL} \and
	M.\,Leone \inst{\DipPD} \and
	P.\,Mastinu \inst{\infnLNL} \and
	R.\,Menegazzo \inst{\infnpd} \and
	G.\,Montagnoli \inst{\infnpd , \DipPD} \and
	C.\,Rossi Alvarez \inst{\infnLNL} \and
	V.\,Rigato \inst{\infnLNL} \and
	O.\,Wieland  \inst{\infnMi} %\and
	%}                     % Do not remove
%
%\thanks{\emph{Present address:} caciolli@pd.infn.it}%
}                     % Do not remove

\institute{
	INFN Sezione di Padova, Padova, Italy % 1
	\and 
	INFN Laboratori Nazionali di Legnaro, Legnaro (Padova), Italy % 2
	\and 
	Dipartimento di Fisica e Astronomia, Universit\`a degli Studi di Padova, Padova, Italy % 3
	\and
	INFN Sezione di Milano, Milano, Italy % 4
	\and
	Nevsehir Haci Bektas University, Science and Art Faculty, Physics Department, Nevsehir, Turkey % 5
} % Do not remove
\date{\today}

\abstract{
The observation of \Al  gives us the proof of  active nucleosynthesis in the Milky Way. However the identification of the main producers of \Al is still a matter of debate. Many sites have been proposed, but our poor knowledge of the nuclear processes involved introduces high uncertainties. In particular, the limited accuracy on the \MgAl reaction cross section has been identified as the main source of nuclear uncertainty in the production of \Al in C/Ne explosive burning in massive stars, which has been suggested to be the main source of \Al in the Galaxy.  We studied this reaction through neutron spectroscopy at the CN Van de Graaff accelerator of the Legnaro National Laboratories. Thanks to this technique we are able to discriminate the ($\alpha$,n) events from  possible contamination  arising from parasitic reactions. In particular, we measured the neutron angular distributions at 5 different beam energies (between 3 and 5 MeV) in the \ang{17.5}-\ang{106} laboratory system angular range.
The presented results  disagree with the assumptions introduced in the analysis of  a previous experiment. %Therefore we show  the need of new experimental efforts to obtain accurate \MgAl total cross section data at astrophysical energies  to re-evaluate the assumptions made in the past. 
 % end of PACS codes
} %end of abstract

%\authorrunning{A. Caciolli {\it et al.}}

%\titlerunning{Not  decided yet}

\maketitle

%===========================
% ==========================
\section{Introduction}
%===========================
%===========================

The abundance of \Al in the Milky Way has been measured  since 1979 by the HEAO C-1  and the SMM/GRS telescopes \cite{SMM,SMM1}. \Al decays into the first excited state of $^{26}$Mg with an half-life of 7.2$\cdot$10$^5$ years.
The interstellar medium is almost transparent to the 1.8 MeV $\gamma$ ray emitted during this decay. The first complete map of \Al in the Milky Way was produced by the CGRO/COMPTEL in 1995 \cite{CGRO} and, recently, INTEGRAL collaborations reported an abundance of \Al in our Galaxy of 2.8$\pm$0.8 solar masses \cite{Diehl-Nature}.

The observation of the \Al decay  provides a ``snapshot" view of the continuing nucleosynthesis in the Galaxy \cite{Prantzos1996}.
The identification of the main sources of $^{26}$Al would have far-reaching implications for many astrophysical issues, ranging from the chemical evolution of the Galaxy to the formation and early evolution of the Solar System. As a matter of fact, \Al is co-rotating with the Galaxy and it can be used to trace the kinematics of cumulative massive-star and supernova ejecta independent of the uncertain gas parameters over million-year time scales \cite{Kretschmer2013}.

The \Al has been observed predominantly in the galactic plane, a star-forming region. This suggests that the \Al is produced by  stars with a mass higher than 8 solar masses, especially during explosive burning \cite{Limongi&Chieffi}.  However, the origin of \Al still remains controversial.

The correct prediction of its abundance could also be a constraint in the determination of the  frequency of core collapse supernova events (that is, type I b/c and type II) \cite{Diehl-Nature}.

In addition, an excess of $^{26}$Mg was found in meteorites indicating the presence of \Al in the hot disk-accretion phase of the presolar nebula \cite{MacPherson1995,Young2005} and in the early phases of the Solar System formation. 
The presence of \Al in the early stages of the Solar System could be associated with a nearby supernova event or with an Asymptotic Giant Branch star  injecting  fresh nucleosynthesis isotopes in the magnetically active early Sun~\cite{Lee1998}.

The \Al is produced in the Mg-Al cycle by the well known $^{25}$Mg(p,$\gamma$)$^{26}$Al reaction \cite{Limata2010,Strieder2012}. This  reaction has been precisely studied by the LUNA experiment \cite{Broggini-review} thanks to its underground location \cite{Bemmerer2005,Caciolli2009}.
While  $^{25}$Mg(p,$\gamma$)$^{26}$Al  produces $^{26}$Al, there are also reactions involved in the destruction of this isotope and its seed,  $^{25}$Mg:  $^{26}$Al(p,$\gamma$)$^{27}$Si, $^{26}$Al(n,$\alpha$)$^{23}$Na, $^{26}$Al(n,p)$^{26}$Mg, and  $^{25}$Mg($\alpha$,n)$^{28}$Si. The first reaction has been recently studied \cite{Lotay2009} while on the others there are still some discussions. Their impact on the \Al production during C/Ne explosive burning in massive stars has been recently studied   and detailed comparisons between  data and models have been performed finding strong discrepancies (\cite{Iliadis2011,Oginni2011} and reference therein).
In particular, Iliadis and coworkers claimed that the nuclear contribution to the \Al uncertainty  is dominated by the uncertainty on the \MgAl reaction rate and they underlined the need for new experimental efforts in order to reduce the errors on the determination of the cross section value \cite{Iliadis2011}.
The relevant energy region for the \MgAl in this context is between 1 and 5 MeV. 
At lower energies, the  \MgAl reaction has also a minor influence on the neutron production for s-processes \cite{Hoffman2001}.

At these astrophysical energies the cross section is highly suppressed due to the effect of the Coulomb barrier. In order to extrapolate the data to these energies it is advantageous to transform the cross section into the astrophysical $S(E)$ factor defined by \cite{Iliadis_book}:
%===========================
\begin{equation}\label{eq:sfactor}
\sigma(E) = \frac{S (E)}{E}\exp({-2 \pi \eta}) 
\end{equation}
%===========================
where $2 \pi \eta = 0.989534$ $ Z_1 Z_2 \sqrt{\frac{\mu}{E}}$ is the Sommerfeld parameter, $Z_1$ and $Z_2$ are the atomic numbers of the interacting ions. The reduced mass, $\mu$, is  expressed in a.m.u. and $E$ is expressed in MeV.

In this paper  we propose an experimental approach to the \MgAl cross section measurement based on neutron spectroscopy. The emitted neutrons are identified using a time of flight (TOF) technique. This approach is ``quasi-free'' from beam induced background, but it suffers from low detection efficiency  compared to the thermalizing neutrons approach (see sec. \ref{sec2}),  due to a lower   solid angle coverage. 
In the following section we report an overview of the existing measurements and their discrepancies. In section \ref{sec:setup} we discuss the experimental setup, while in section \ref{sec:analysis} the analysis procedure is described. 
Finally the measured angular distributions are given in section \ref{sec:results}. 
%In particular, our results show that the angular dependence of the cross section has to be measured with fine granularity in an angular region as wide as possible.

%===========================
%===========================
\section{Existing measurements}\label{sec2}
%===========================
%===========================

\begin{figure}[htb]
\centerline{%
\includegraphics[width=\columnwidth]{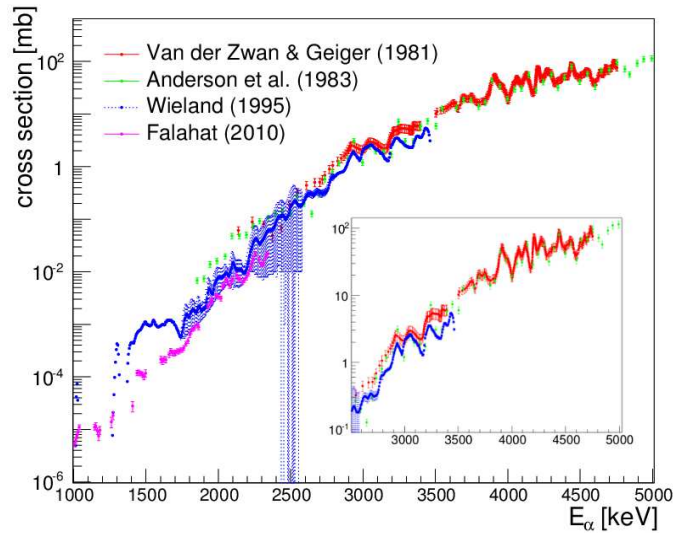}}
\caption{Previous experimental data. The Van Der Zwan and Geiger data are in red circle, Anderson \emph{et al.} data in green open crosses, Wieland data are reported in blue open triangle, and the Falahat data in pink open square. It has to be noted that below 1.7 MeV the Wieland data are only upper limits. NACRE uses only the Wieland data and performs HF calculation at energies above 2.5 MeV. In the inset a zoom of the existing data in the energy range of interest of the present work is shown.}\label{literature_data}
\end{figure}

\begin{figure}[htb]
\centerline{%
\includegraphics[width=7cm]{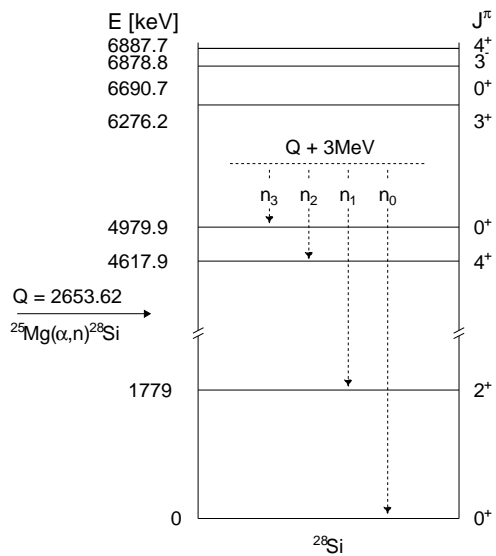}}
\caption{Partial level scheme of $^{28}$Si. The Q-value of \MgAl is reported and the n$_0$ to n$_3$ neutron branchings are schematically shown for $E_\alpha$ =  3 MeV.}\label{levscheme}
\end{figure}

The \MgAl (Q = 2.654 MeV) has been measured in 1981 by Van Der Zwan and Geiger \cite{VanderZwan1981} and then by Anderson and coworkers in 1983 \cite{Anderson1983}. 
In the former work, neutron spectroscopy was performed using 4 Stilbene scintillators placed at \ang{0}, \ang{45}, \ang{90}, and \ang{135} in the laboratory reference frame. The measured differential cross section was fitted with Legendre polynomials from P$_0$ to P$_2$  for emitted neutrons corresponding to the population of ground and 1st excited states of $^{28}$Si. For the population of higher excited states (2nd and 3rd) an isotropic distribution was assumed in \cite{VanderZwan1981}
(see in fig. \ref{levscheme} the $^{28}$Si level scheme). 
The cross section as a function of the incident beam energy in the lab frame is reported in fig. \ref{literature_data} as red points.

Anderson and collaborators measured both the neutrons and the $\gamma$ rays produced by the reaction. The neutron detector used was made of BF$_3$ tubes with an active length of 40 cm embedded in a 50 cm large paraffin cube  with the target in the centre. For the $\gamma$ rays a Ge(Li) detector was used. The deduced cross section is reported as green points in fig. \ref{literature_data}. Statistical model calculations were used to predict the branching, the energy and the angular distribution of  emitted neutrons. According to them, the beam induced background,  due to ($\alpha$,n) and ($\alpha$,n$\gamma$) reactions on $^{13}$C, $^{18}$O, and $^{19}$F contaminants in the target, was not clearly resolved. 
%They claim a beam induced background not clearly resolved due to ($\alpha$,n) and ($\alpha$,n$\gamma$) reactions on $^{13}$C, $^{18}$O, and $^{19}$F contaminants in the target. 
%%%
The same beam induced reactions on contaminants  were observed by Wieland \cite{Wieland} who measured the \MgAl cross section down to 0.874 MeV with a 4$\pi$ Helium counter detector. The results are reported in fig. \ref{literature_data} as blue points.
The NACRE database \cite{NACRE} reports the data from \cite{VanderZwan1981}, \cite{Anderson1983} and \cite{Wieland} but only the Wieland's data are used to evaluate the cross section at energies between 0.8 and 2.5 MeV. 

%===========================
%===========================
In the energy range above 2.5 MeV NACRE collaboration determines the S-factor  by using the Hauser-Feshbach (HF) calculations. In the energy range below the lowest data point reported in Wieland's thesis the S-factor has been considered to be constant. 
Iliadis and coworkers \cite{Iliadis2011} report a discussion on the data choice by NACRE emphasising the differences between the adopted cross section  and  results published in \cite{VanderZwan1981} and \cite{Anderson1983}. 
Finally,  they strongly pointed out the need for new measurements in the energy range above 1.5 MeV.
In 2010 new experimental results on \MgAl below 2.5 MeV have been reported by Falahat \cite{Falahat2010}.
As shown in fig. \ref{literature_data}, this data  shows a difference of a factor 3 with respect to Wieland data although they are still in agreement within the large error bars of \cite{Wieland}. The data of \cite{Falahat2010} are not consistent with \cite{VanderZwan1981} and \cite{Anderson1983}.
The experimental setup described in \cite{Falahat2010} is based on the same concept used by Wieland and so it is similarly unable to distinguish neutrons produced by the \MgAl reaction from those emitted by reactions on  contaminants.
These issues strongly suggest the need for a new cross section measurement for this reaction.
A new study of this reaction was performed in 2013 \cite{Negret2013} in an energy range above 5.6 MeV. 
%They do not measure the emitted neutrons, but the $\gamma$ rays produced by the decay of $^{28}$Si excited states.
In this measurement, only the $\gamma$ rays produced by the the decay of $^{28}$Si excited states were detected.

% ========================================================
%===========================
\section{Experimental Setup}\label{sec:setup}
%===========================
% ========================================================
The experiment was performed at the \ang{0} beam-line of the CN Van de Graaff accelerator at  Legnaro National Laboratories (INFN-LNL). A pulsed $^4$He$^+$ beam was used (I$\approx$ 200 pnA, 3 MHz repetition rate, bunch width $<$2 ns). The measurements were performed at 5 different energies from 3 MeV to 5 MeV.

%===========================
\begin{figure}[htb]
\centerline{%
\includegraphics[width=\columnwidth]{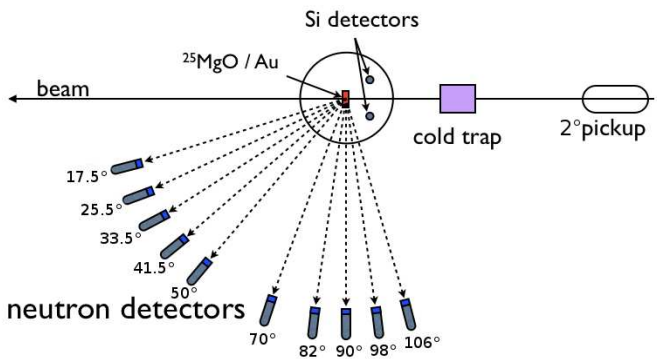}}
\caption{Schematic picture of the setup. See text for details.}\label{fig:setup}
\end{figure}
%===========================

Neutrons were detected by using 10 BC501 liquid scintillators from the RIPEN array \cite{RIPEN}, as already done in a previous experiment using a similar setup~\cite{BETABEAM}. Each detector has an active cell of 12.5 cm diameter and 12.5 cm length. In order to reduce the background induced by low-energy $\gamma$ rays each cell was surrounded with a 5 mm lead shielding. The center of each cell was placed at a distance of (206.3$\pm$0.5) cm from the target position covering an angular range from \ang{17.5} to \ang{106} with respect to the beam direction (see fig. \ref{fig:setup}). Each detector covers a 3 msr solid angle. The time of flight technique was used to determine the neutron energy on an event by event basis, while pulse shape analysis provided discrimination between n and $\gamma$.

Two inductive pickup devices placed 9.40 m apart provided the reference signal needed for the time of flight measurement. The first pickup was placed at 10.90 m before the target position and the second one at 1.5 m.
Moreover, the time of flight between the two pickups allowed to measure the beam energy. The distance between the two pickups was precisely determined through the $^{16}$O($\alpha$,$\alpha$)$^{16}$O resonance at 3.045 MeV  \cite{Cameron-1953} which is known to have a FWHM lower than 10 keV. With this method a 0.7\% uncertainty on the energy was achieved for the 3 MeV $\alpha$-beam.

Two collimated Silicon detectors (1 mm  diameter) were placed inside the chamber at a distance of 7 cm from the target and at $\pm$\ang{150} with respect to the beam direction.
The silicon detectors have been used for two  purposes: (i) to count the backward scattered $\alpha$ ions for cross section normalisation, (ii) to monitor the target integrity and the possible presence of contaminants.

The target holder was made of copper and was cooled with water at a temperature of \ang{14}C in order to limit the target deterioration. 
The beam focusing was checked regularly using a copper frame with a 7 mm diameter hole (1 mm less than the target diameter) minimising the beam current on the copper frame.

From previous experiments \cite{Anderson1983,Wieland,Falahat2010}, we know that carbon and fluorine contaminants should be the main sources of background.
As discussed in the next section, the decision of performing neutron spectroscopy is based mainly on the possibility to identify   neutrons produced by the \MgAl with respect to other ($\alpha$,n) reactions. In addition, the background problem was addressed by installing two turbomolecular pumps and a cold trap as close as possible to the scattering chamber. This allowed us to keep the vacuum in the scattering chamber always below 8$\cdot 10^{-7}$ mbar. In particular, the cold trap was necessary to reduce the carbon build-up on the target surface. It was made of a copper tube (2 cm diameter 20 cm length) placed at 1 m  from the target and cooled with liquid nitrogen.

%===========================
%===========================
\subsection{Targets preparation and analysis}\label{sec:target}
%===========================
%===========================
Targets were made of MgO evaporated on gold. The Mg was 95.75\% enriched in $^{25}$Mg. The nominal target thickness was 70 $\mu$g/cm$^2$ while the gold backing was 1 mg/cm$^2$ thick. 
The oxide prevents the magnesium from being further oxidized and keeps the $^{18}$O background constant during the experiment. Moreover, we stress the fact that the isotopic abundance of $^{18}$O is only 0.2\% and that the neutrons produced by the $^{18}$O($\alpha$,n)$^{21}$Ne reaction (Q = -0.697	MeV) have energies that can be easily discriminated by TOF. 
The $^{21}$Ne has three excited states that could be populated in the energy range explored by the present work: 0.35 MeV, 1.75 MeV and 2.19 MeV. Due to the negative Q-value of this reaction the emitted neutrons have considerably lower energies with respect to the \MgAl ones and they are found in a region of the TOF spectrum clearly separated from our region of interest. 
Beam induced background from   $^{17}$O($\alpha$,n)$^{20}$Ne reaction (Q = 0.587 MeV) has been investigated and it can be considered negligible. In particular,  its reaction cross section is lower than the $^{18}$O($\alpha$,n)$^{21}$Ne and $^{17}$O abundance in natural oxygen is a factor of 5 lower than $^{18}$O.

The targets were analysed using Rutherford Backscattering Spectroscopy (RBS)  at the AN2000 accelerator with the setup already used in \cite{CaciolliTaO}. Negligible carbon contamination was observed on the target during this analysis. 
Elastic cross section measurements are only available for natural magnesium and there is no data for the $^{25}$Mg isotope. This prevents the possibility to determine the target thickness and Mg:O stoichiometry with this method.
We opted to use the nominal value given by the producer for the data analysis and cross section calculation to have an internal normalisation between all measurements, but we decided to give the differential cross section value in arbitrary units as specified in the following sections.
It has to be noted that the target thickness  does not affect the shape of the experimental angular distribution, but only its absolute scale. 

%===========================
%===========================
\subsection{Data Acquisition}
%===========================
%===========================

The experimental setup was completely equipped with digital electronics. An updated version of the acquisition system reported in~\cite{BETABEAM} was used. The signal from the BC501 detectors was acquired by three CAEN V1720 (250 MS/s, 12 bit, 2 V$_{pp}$ dynamic range), while the signal from the silicon detectors by one CAEN V1724 (100 MS/s, 14 bit, 2.25 V$_{pp}$ dynamic range) digitisers. The clock was synchronised between the different boards and it is natively synchronised between the different channels in each board. 

The trigger requests for each digitiser are collected and evaluated according to a software programmable configuration matrix. The global trigger signal is then sent back to the digitisers. Two independent triggers have been used during the experiments: the OR of the 10 liquid scintillators and the OR of the two silicon detectors.

Three different kind of informations are expected to be obtained processing the scintillator signals: the energy release of the impinging radiation, its time of fight and the pulse shape discrimination between neutrons and $\gamma$ rays. After proper baseline subtraction, the energy release is estimated via signal integration. The time of flight is obtained as the difference between a digital constant fraction  filter   applied to the BC501 signal shapes and the  reference from the second pickup signal. For the details on the algorithm used see~\cite{ZCO}. The n/$\gamma$ discrimination is achieved using a digital implementation of the Zero-Crossing method~\cite{ZCO}, the result is a mono dimensional variable (ZCO) that can be directly plotted against time of flight or deposited energy to distinguish neutrons from $\gamma$ rays as described in details in the next section. The energy measured by the silicon detectors has been evaluated as the maximum of the amplified signal.

%===========================
%===========================
\section{Data Analysis}\label{sec:analysis}
%===========================
%===========================

%===========================
%===========================

\begin{figure}[htb]
\centerline{%
\includegraphics[width=\columnwidth]{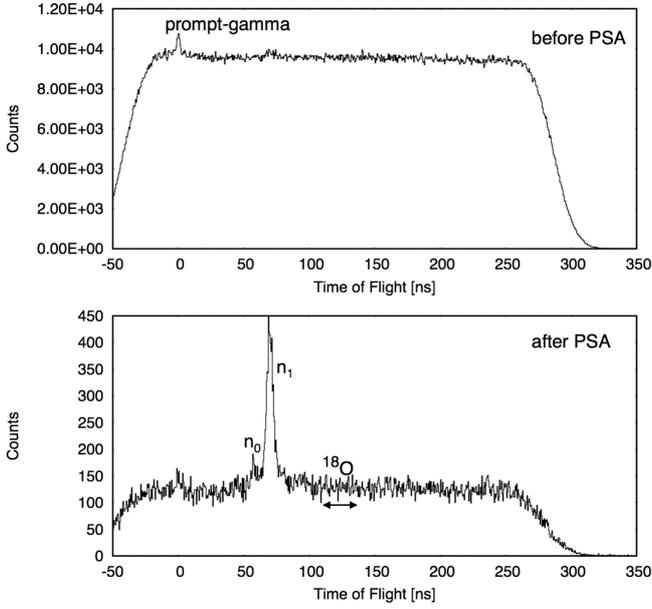}}
 \caption{Trigger rate of the BC501 as fuction of the TOF. The spectrum was acquired at \ang{25.5} with a $\alpha$-energy of 3 MeV. Upper panel: no PSA is used to reconstruct the spectrum. Lower panel: the events are reconstructed selecting a defined region of the ZCOvsEnergy spectrum (see fig. \ref{fig:2Dplot}). The n$_0$ and n$_1$ peaks and the region where we expect the peaks from $^{18}$O contaminant are shown in lower panel.} \label{fig:spectrum5MeV}
 \end{figure}

\begin{figure}
\centerline{%
\includegraphics[width=\columnwidth]{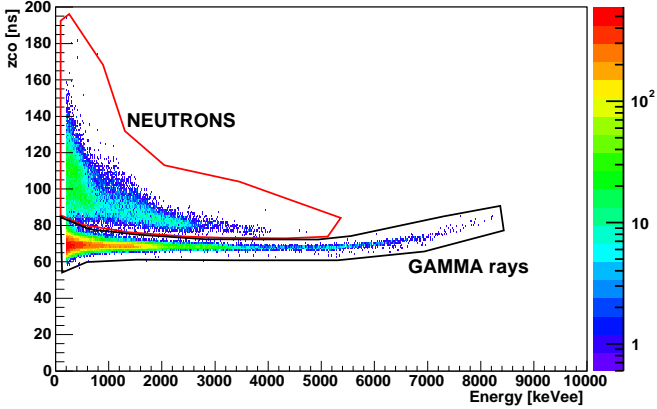}}
 \caption{The ZCO versus deposited energy in the scintillator detector is shown in the same experimental condition of fig.\ref{fig:spectrum5MeV}. Two regions are clearly visible. The neutron events are highlighted).} \label{fig:2Dplot}
 \end{figure}

The experimental yields are reconstructed from the TOF spectra.
The prompt-$\gamma$ peak is used as a reference to measure the absolute value of the time of flight for the neutrons. The neutron energy is then uniquely determined by the measured TOF. 
For a defined beam energy the emitted neutrons have kinetic energy in the center of mass equal to $E_{CM} + Q - E_X$, where $E_{CM}$ is the energy in the center of mass system and $E_X$ is the energy of the populated $^{28}$Si excited level. 
Due to the low \MgAl cross section, the TOF spectrum is dominated by the uncorrelated $\gamma$ rays produced by the environmental radioactivity (upper panel of fig. \ref{fig:spectrum5MeV}). 
Plotting  ZCO versus the deposited energy in the detector, two regions are clearly seen as shown in fig. \ref{fig:2Dplot}. The lower region is related to the signal produced by  $\gamma$ rays, whereas the upper one  is related to neutron like events.
The $\gamma$ and neutron regions are well separated and therefore it is possible to reduce the $\gamma$ background in the TOF spectra  selecting only  neutron events as illustrated in the lower panel of fig. \ref{fig:spectrum5MeV}. 
After this selection the two neutron peaks related to the population of the ground (n$_0$) and first excited (n$_1$) states of $^{28}$Si emerge from the TOF spectrum as shown in fig.\ref{fig:spectrum5MeV}.
The still present flat structure is most probably  due to uncorrelated neutron events produced by the neutron component of the cosmic ray flux. 
Another small component,  due to residual $\gamma$ rays, could be present due to the small overlap between the $\gamma$ and neutron regions as can be seen in fig.\ref{fig:2Dplot}. 
%===========================
%===========================
The experimental yield at each angle has been calculated integrating the peak areas, $A_n$, and using the following equation:
%===========================
\begin{equation}\label{eq:YieldExp}
\frac{\mbox{d}Y}{\mbox{d}\Omega} = \frac{A_n}{\eta \Omega N_\alpha}
\end{equation}
%===========================
where $\Omega$ is the solid angle covered by each scintillator, $N_\alpha$ is the integrated beam  charge expressed as the number of  $\alpha$ particles, and $\eta$ is the intrinsic efficiency for neutron detection. The intrinsic efficiency has been determined by Monte Carlo simulations already tested with the RIPEN detectors \cite{BETABEAM}. 
%===========================
%===========================
The efficiency curve  
depends on the detectors threshold. 
In the present experiment the threshold was set at 200 keVee for all detectors in order to cut the electronic noise. This correspond to a minimum detectable neutron energy of about 1 MeV. 

The integrated beam  charge is evaluated  by integrating the gold peak in the RBS spectra, acquired with the silicon detectors. That area is produced by the Rutherford backscattering of $\alpha$ particles on the gold target backing. The final value for the integrated charge was determined from the average of the two symmetric Silicon detectors.

The experimental yield determined in eq. (\ref{eq:YieldExp}) is then compared with a yield calculated using the  S-factor reported in the NACRE database \cite{NACRE}. The S-factor can be assumed  to be constant  over the energy range explored by the target thickness, therefore the cross section can be calculated from eq. (\ref{eq:sfactor}) and the yield calculated  by using the following formula:
%===========================
\begin{equation}\label{eq:YieldCalc}
\left(\frac{\mbox{d}Y}{\mbox{d}\Omega}\right)_{Calc} = \int^{E_\alpha}_{E\alpha - \Delta E}  \left(\frac{\mbox{d}\sigma}{\mbox{d}\Omega}\right)_{Th} \frac{1}{\epsilon _{eff}}\mbox{d}E
\end{equation}
%===========================
where $\Delta E$ is the $\alpha$ energy loss in the target and $\epsilon _{eff}$ is the effective stopping power which is expressed as \cite{Iliadis_book}:
%===========================
\begin{equation}\label{eq:eff_stop}
\epsilon _{eff} = (\epsilon_{Mg} + \epsilon_{O}) / 0.9575.
\end{equation}
%===========================
The stopping power values are taken from SRIM \cite{SRIM}.

The experimental cross section is then determined from the  ratio of eq. (\ref{eq:YieldExp}) and eq. (\ref{eq:YieldCalc}):
%===========================
\begin{equation}
\left(\frac{\mbox{d}\sigma}{\mbox{d}\Omega}\right)_{Exp} = \left(\frac{\mbox{d}\sigma}{\mbox{d}\Omega}\right)_{Th} \cdot \frac{\left(\frac{\mbox{d}Y}{\mbox{d}\Omega}\right)_{Exp}}{\left(\frac{\mbox{d}Y}{\mbox{d}\Omega}\right)_{Calc}}.
\end{equation}
%===========================

Due to the uncertainties on the target composition and thickness in section \ref{sec:target}  the angular distribution is reported in the following only as relative probability. We would like to stress that the shape of the distribution is not affected by these uncertainties.
%we decided to give the angular distribution obtained with this procedure only as a relative probability and not as an absolute value. Still the information on the distribution shape is not affected by these uncertainties.

%===========================
%===========================
\section{Results}\label{sec:results}
%===========================
%===========================

\begin{figure}
\centerline{%
\includegraphics[width=\columnwidth]{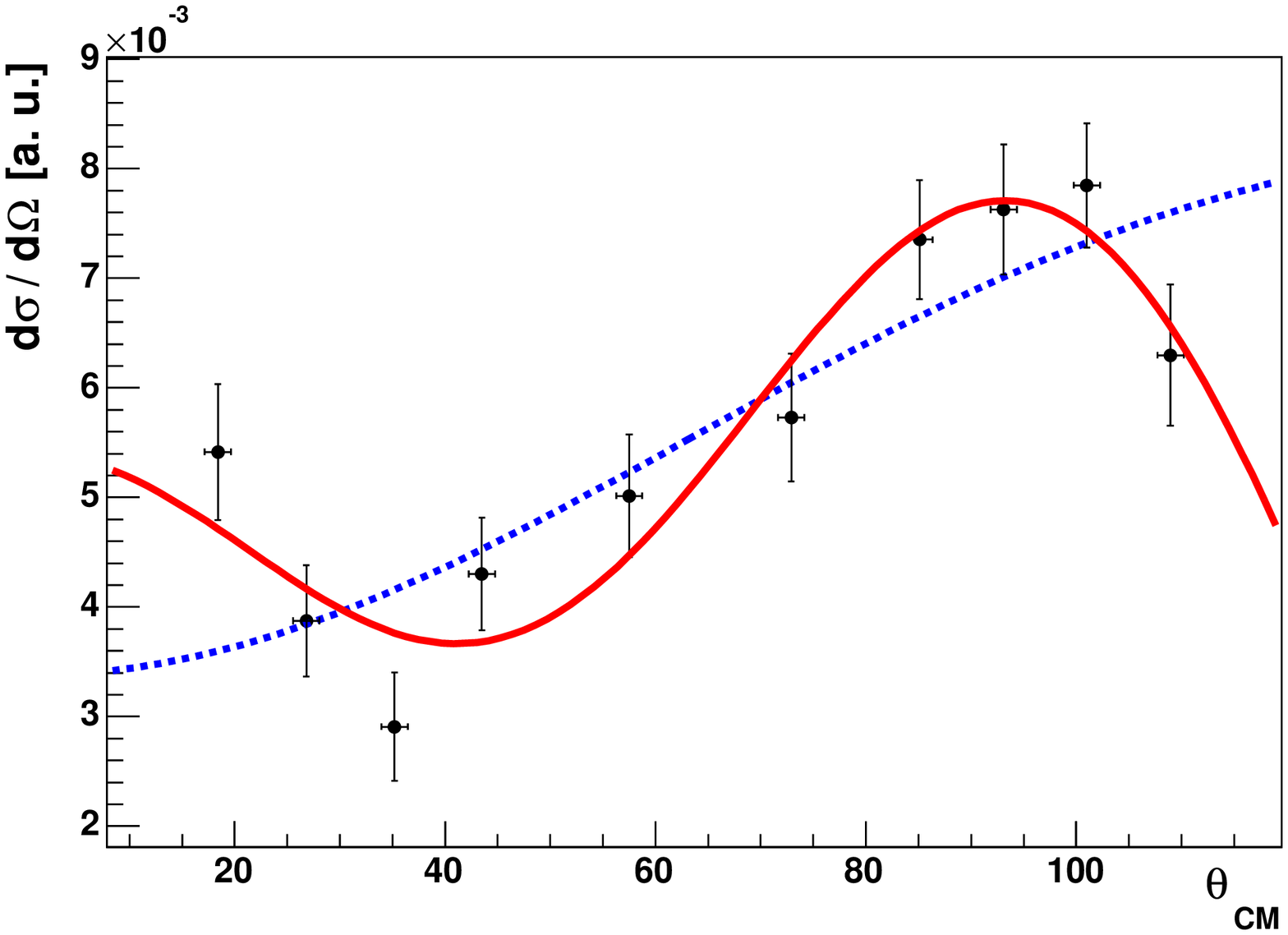}}
 \caption{The angular distribution of n$_0$ at  the $\alpha$-beam energy of 3.045 MeV are reported in black. The fit using from P$_0$ to P$_2$ Legendre polynomials is in blue dashed line, while the fit made including the Legendre polynomials until P4 is reported in red line.} \label{fig:crosssection3_n0}
\end{figure}

\begin{figure}
\centerline{%
\includegraphics[width=\columnwidth]{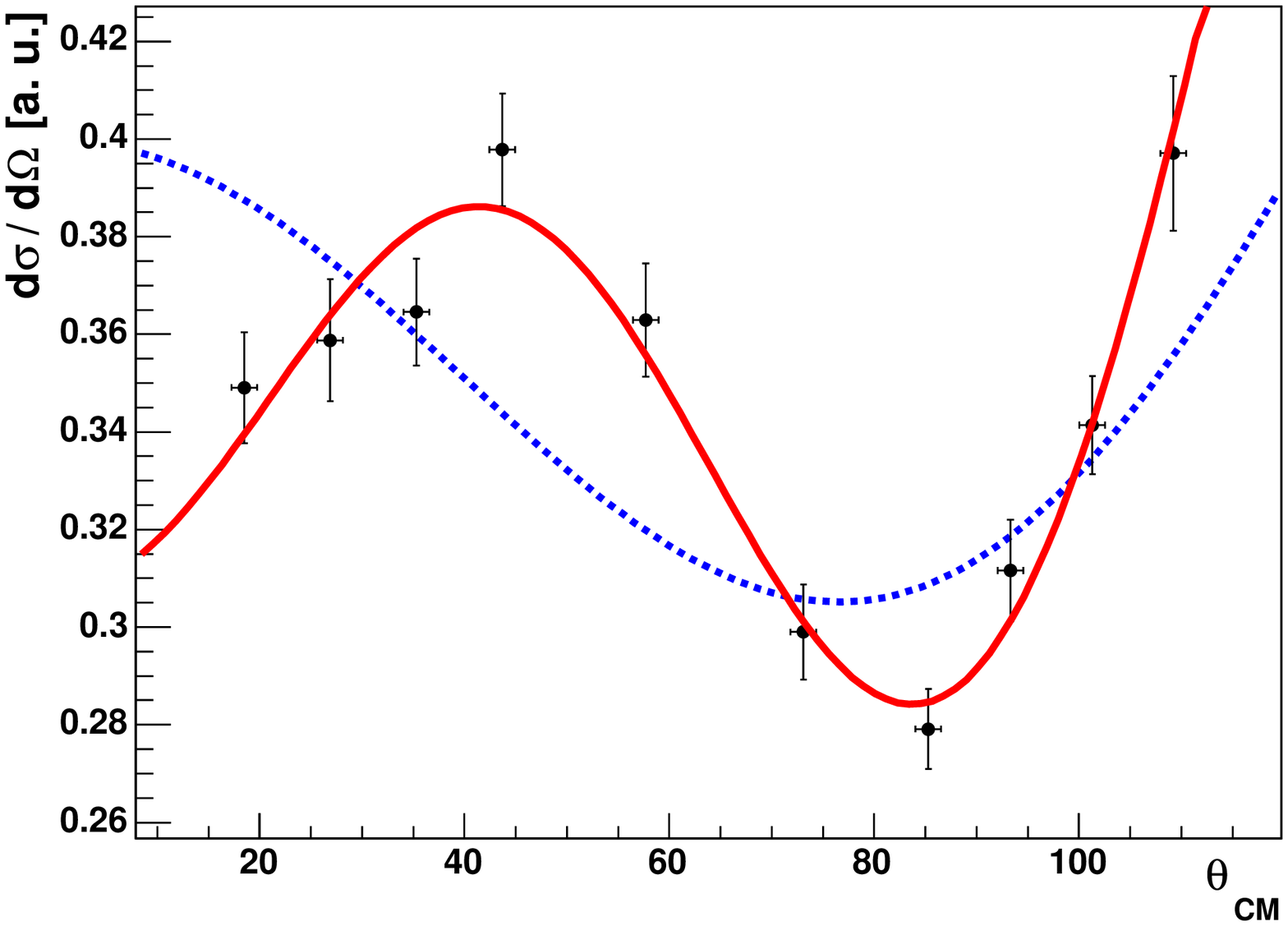}}
 \caption{The angular distribution of n$_1$ at  the $\alpha$-beam energy of 4.3 MeV are reported in black. The fit using from P$_0$ to P$_2$ Legendre polynomials is in green  dashed line, while the fit made including the Legendre polynomials until P$_4$ is reported in red line.} \label{fig:crosssection4_5}
\end{figure}

\begin{figure}
\centerline{%
\includegraphics[width=\columnwidth]{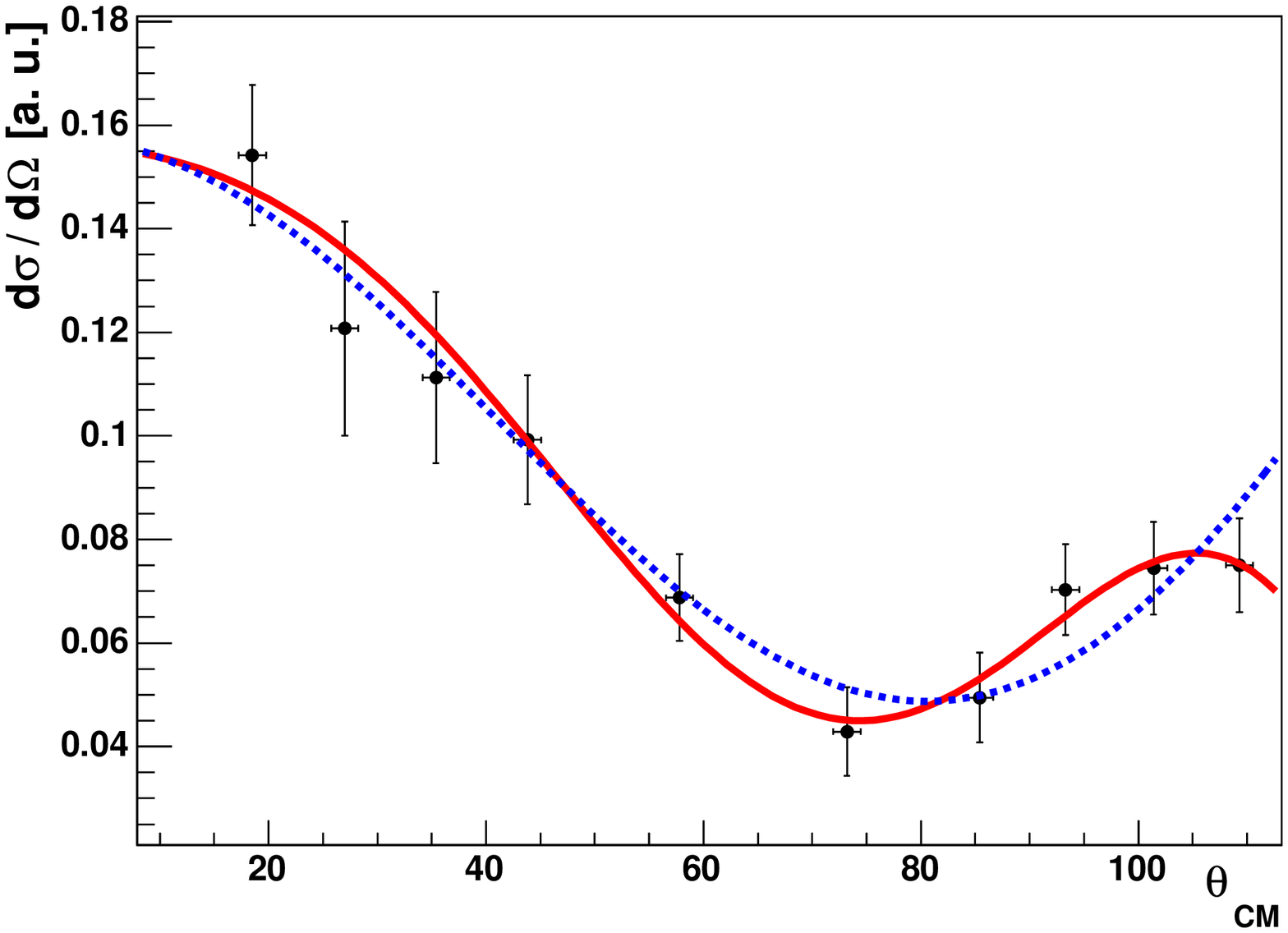}}
 \caption{The angular distribution of n$_0$ at  the $\alpha$-beam energy of 4.95 MeV are reported in black. The fit using from P$_0$ to P$_2$ Legendre polynomials is in blue dashed line, while the fit made including the Legendre polynomials until P$_4$ is reported in red line.} \label{fig:crosssection5_n0}
 \end{figure}

\begin{figure}
\centerline{%
\includegraphics[width=\columnwidth]{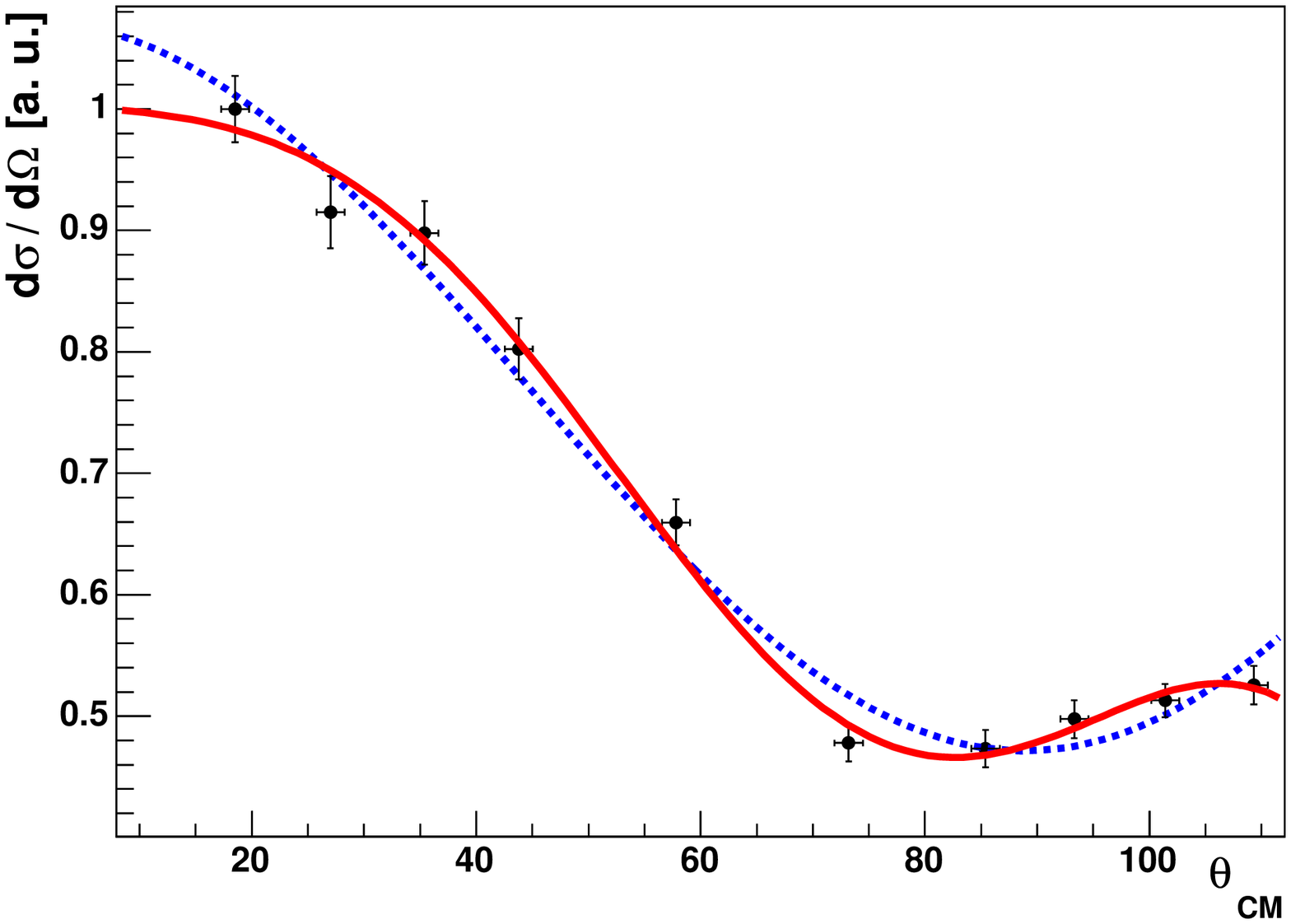}}
 \caption{The angular distribution of n$_1$ at  the $\alpha$-beam energy of 4.95 MeV are reported in black. The fit using from P$_0$ to P$_2$ Legendre polynomials is in blue dashed line, while the fit made including the Legendre polynomials until P$_4$ is reported in red line.} \label{fig:crosssection5_n1}
 \end{figure}

The  angular distributions measured in this experiment are reported in Tables \ref{table1} - \ref{table5}. 
Some examples of these distributions are shown in     figs \ref{fig:crosssection3_n0} - \ref{fig:crosssection5_n1}. 
Only statistical errors are considered and the errors on the angles are assumed equal to the angular dimension of the RIPEN detectors ($\sim$1.75$^{\circ}$). 
The relative differential cross sections  are always normalised to the differential cross section at \ang{18.5} relative to the population of the $^{28}$Si first excited state  (n$_1$) at $E_\alpha$ = 4.95 MeV (see Table \ref{table5}).
%==========================================
\begin{table*}
\centering
\caption{Measured angular distributions  for the n$_0$ and n$_1$ transitions at $E_\alpha$ = 3.045 MeV}\label{table1}
\begin{tabular}[!h]{ccc}
\hline \hline
$\theta_{CM}$[degrees]	&	n$_0$						&	n$_1$						\\
18.4	&	(	5.4	$\pm$	0.6	)	$\cdot 10^{-3}$	&	(	4.65	$\pm$	0.08	)	$\cdot 10^{-2}$	\\
26.8	&	(	3.9	$\pm$	0.5	)	$\cdot 10^{-3}$	&	(	4.16	$\pm$	0.08	)	$\cdot 10^{-2}$	\\
35.2	&	(	2.9	$\pm$	0.5	)	$\cdot 10^{-3}$	&	(	3.70	$\pm$	0.03	)	$\cdot 10^{-2}$	\\
43.5	&	(	4.3	$\pm$	0.5	)	$\cdot 10^{-3}$	&	(	3.58	$\pm$	0.03	)	$\cdot 10^{-2}$	\\
57.5	&	(	5.0	$\pm$	0.6	)	$\cdot 10^{-3}$	&	(	3.24	$\pm$	0.03	)	$\cdot 10^{-2}$	\\
72.9	&	(	5.7	$\pm$	0.6	)	$\cdot 10^{-3}$	&	(	3.01	$\pm$	0.05	)	$\cdot 10^{-2}$	\\
85.1	&	(	7.4	$\pm$	0.5	)	$\cdot 10^{-3}$	&	(	2.95	$\pm$	0.03	)	$\cdot 10^{-2}$	\\
93.1	&	(	7.6	$\pm$	0.6	)	$\cdot 10^{-3}$	&	(	3.15	$\pm$	0.03	)	$\cdot 10^{-2}$	\\
101.0	&	(	7.8	$\pm$	0.6	)	$\cdot 10^{-3}$	&	(	3.42	$\pm$	0.03	)	$\cdot 10^{-2}$	\\
109.0	&	(	6.3	$\pm$	0.6	)	$\cdot 10^{-3}$	&	(	3.26	$\pm$	0.03	)	$\cdot 10^{-2}$	\\
\hline												
\hline																			
\end{tabular}
\end{table*}
%==========================================			
%==========================================
\begin{table*}
\centering
\caption{Measured angular distributions for the n$_0$, n$_1$ and n$_2$ transitions at $E_\alpha$ = 3.49 MeV}\label{table2}
\begin{tabular}[!h]{cccc}
\hline \hline
$\theta_{CM}$[degrees]	&	n$_0$						&	n$_1$						&	n$_2$						\\
18.5	&	(	8.9	$\pm$	0.5	)	$\cdot 10^{-3}$	&	(	8.1	$\pm$	0.2	)	$\cdot 10^{-2}$	&	(	12.8	$\pm$	1.1	)	$\cdot 10^{-3}$	\\
26.9	&	(	8.1	$\pm$	0.9	)	$\cdot 10^{-3}$	&	(	6.8	$\pm$	0.2	)	$\cdot 10^{-2}$	&	(	10.2	$\pm$	1.1	)	$\cdot 10^{-3}$	\\
35.3	&	(	6.3	$\pm$	0.4	)	$\cdot 10^{-3}$	&	(	6.20	$\pm$	0.09	)	$\cdot 10^{-2}$	&	(	7.9	$\pm$	1.2	)	$\cdot 10^{-3}$	\\
43.6	&	(	6.8	$\pm$	0.4	)	$\cdot 10^{-3}$	&	(	5.93	$\pm$	0.09	)	$\cdot 10^{-2}$	&	(	3.9	$\pm$	1.3	)	$\cdot 10^{-3}$	\\
57.6	&	(	6.3	$\pm$	0.4	)	$\cdot 10^{-3}$	&	(	4.30	$\pm$	0.07	)	$\cdot 10^{-2}$	&	(	7.0	$\pm$	1.8	)	$\cdot 10^{-3}$	\\
73	&	(	5.8	$\pm$	0.4	)	$\cdot 10^{-3}$	&	(	3.46	$\pm$	0.13	)	$\cdot 10^{-2}$	&	(	6	$\pm$	4	)	$\cdot 10^{-3}$	\\
85.2	&	(	6.4	$\pm$	0.4	)	$\cdot 10^{-3}$	&	(	3.22	$\pm$	0.08	)	$\cdot 10^{-2}$	&		below detection threshold				\\
93.2	&	(	6.3	$\pm$	0.4	)	$\cdot 10^{-3}$	&	(	3.86	$\pm$	0.07	)	$\cdot 10^{-2}$	&		below detection threshold			\\
101.2	&	(	7.5	$\pm$	0.4	)	$\cdot 10^{-3}$	&	(	4.32	$\pm$	0.09	)	$\cdot 10^{-2}$	&		below detection threshold			\\
109.1	&	(	7.1	$\pm$	0.4	)	$\cdot 10^{-3}$	&	(	4.19	$\pm$	0.07	)	$\cdot 10^{-2}$	&		below detection threshold			\\									
\hline
\hline																			
\end{tabular}
\end{table*}
%=========================================
%==========================================
\begin{table*}
\centering
\caption{Measured angular distributions for the n$_0$, n$_1$ and n$_2$ transitions at $E_\alpha$ = 3.95 MeV}\label{table3}
\begin{tabular}[!h]{cccc}
\hline \hline
$\theta_{CM}$[degrees]	&	n$_0$						&	n$_1$						&	n$_2$						\\
18.5	&	(	10.1	$\pm$	1.0	)	$\cdot 10^{-2}$	&	(	2.50	$\pm$	0.15	)	$\cdot 10^{-1}$	&	(	1.60	$\pm$	0.06	)	$\cdot 10^{-1}$	\\
26.9	&	(	8.0	$\pm$	0.7	)	$\cdot 10^{-2}$	&	(	2.36	$\pm$	0.12	)	$\cdot 10^{-1}$	&	(	1.20	$\pm$	0.06	)	$\cdot 10^{-1}$	\\
35.3	&	(	7.2	$\pm$	0.7	)	$\cdot 10^{-2}$	&	(	2.25	$\pm$	0.12	)	$\cdot 10^{-1}$	&	(	1.06	$\pm$	0.05	)	$\cdot 10^{-1}$	\\
43.7	&	(	6.5	$\pm$	0.8	)	$\cdot 10^{-2}$	&	(	2.31	$\pm$	0.14	)	$\cdot 10^{-1}$	&	(	0.91	$\pm$	0.07	)	$\cdot 10^{-1}$	\\
57.7	&	(	6.9	$\pm$	0.8	)	$\cdot 10^{-2}$	&	(	2.08	$\pm$	0.13	)	$\cdot 10^{-1}$	&	(	0.49	$\pm$	0.04	)	$\cdot 10^{-1}$	\\
73.1	&	(	7.0	$\pm$	0.6	)	$\cdot 10^{-2}$	&	(	1.61	$\pm$	0.08	)	$\cdot 10^{-1}$	&	(	0.78	$\pm$	0.06	)	$\cdot 10^{-1}$	\\
85.2	&	(	6.3	$\pm$	0.6	)	$\cdot 10^{-2}$	&	(	1.66	$\pm$	0.09	)	$\cdot 10^{-1}$	&	(	0.41	$\pm$	0.05	)	$\cdot 10^{-1}$	\\
93.3	&	(	6.7	$\pm$	0.7	)	$\cdot 10^{-2}$	&	(	1.89	$\pm$	0.10	)	$\cdot 10^{-1}$	&	(	0.26	$\pm$	0.05	)	$\cdot 10^{-1}$	\\
101.2	&	(	7.5	$\pm$	0.7	)	$\cdot 10^{-2}$	&	(	1.66	$\pm$	0.10	)	$\cdot 10^{-1}$	&		below detection threshold					\\
109.1	&	(	7.0	$\pm$	0.6	)	$\cdot 10^{-2}$	&	(	1.98	$\pm$	0.10	)	$\cdot 10^{-1}$	&		below detection threshold					\\								
\hline
\hline																			
\end{tabular}
\end{table*}
%=========================================
%==========================================
\begin{table*}
\centering
\caption{Measured angular distributions for the n$_0$, n$_1$, n$_2$, and n$_3$ transitions at $E_\alpha$ = 4.3 MeV}\label{table4}
\begin{tabular}[!h]{ccccc}
\hline \hline
$\theta_{CM}$[degrees]	&	n$_0$						&	n$_1$						&	n$_2$						&	n$_3$						\\
18.5	&	(	6.9	$\pm$	0.6	)	$\cdot 10^{-2}$	&	(	3.49	$\pm$	0.11	)	$\cdot 10^{-1}$	&	(	1.97	$\pm$	0.07	)	$\cdot 10^{-1}$	&	(	6.6	$\pm$	0.5	)	$\cdot 10^{-2}$	\\
26.9	&	(	6.3	$\pm$	0.6	)	$\cdot 10^{-2}$	&	(	3.59	$\pm$	0.13	)	$\cdot 10^{-1}$	&	(	2.17	$\pm$	0.07	)	$\cdot 10^{-1}$	&	(	5.5	$\pm$	0.5	)	$\cdot 10^{-2}$	\\
35.3	&	(	6.5	$\pm$	0.5	)	$\cdot 10^{-2}$	&	(	3.64	$\pm$	0.11	)	$\cdot 10^{-1}$	&	(	2.38	$\pm$	0.08	)	$\cdot 10^{-1}$	&	(	6.7	$\pm$	0.6	)	$\cdot 10^{-2}$	\\
43.7	&	(	6.7	$\pm$	0.5	)	$\cdot 10^{-2}$	&	(	3.98	$\pm$	0.12	)	$\cdot 10^{-1}$	&	(	2.59	$\pm$	0.10	)	$\cdot 10^{-1}$	&	(	5.7	$\pm$	0.7	)	$\cdot 10^{-2}$	\\
57.7	&	(	10.0	$\pm$	0.7	)	$\cdot 10^{-2}$	&	(	3.63	$\pm$	0.12	)	$\cdot 10^{-1}$	&	(	2.74	$\pm$	0.08	)	$\cdot 10^{-1}$	&	(	3.2	$\pm$	0.4	)	$\cdot 10^{-2}$	\\
73.1	&	(	11.0	$\pm$	0.6	)	$\cdot 10^{-2}$	&	(	2.99	$\pm$	0.10	)	$\cdot 10^{-1}$	&	(	3.03	$\pm$	0.10	)	$\cdot 10^{-1}$	&	(	4.4	$\pm$	0.6	)	$\cdot 10^{-2}$	\\
85.3	&	(	12.2	$\pm$	0.6	)	$\cdot 10^{-2}$	&	(	2.79	$\pm$	0.08	)	$\cdot 10^{-1}$	&	(	2.95	$\pm$	0.10	)	$\cdot 10^{-1}$	&	(	2.1	$\pm$	0.6	)	$\cdot 10^{-2}$	\\
93.3	&	(	15.8	$\pm$	0.8	)	$\cdot 10^{-2}$	&	(	3.12	$\pm$	0.10	)	$\cdot 10^{-1}$	&	(	2.86	$\pm$	0.09	)	$\cdot 10^{-1}$	&	(	2.6	$\pm$	0.5	)	$\cdot 10^{-2}$	\\
101.3	&	(	17.3	$\pm$	0.8	)	$\cdot 10^{-2}$	&	(	3.41	$\pm$	0.10	)	$\cdot 10^{-1}$	&	(	2.30	$\pm$	0.08	)	$\cdot 10^{-1}$	&	(	1.1	$\pm$	0.5	)	$\cdot 10^{-2}$	\\
109.2	&	(	17.1	$\pm$	1.1	)	$\cdot 10^{-2}$	&	(	3.97	$\pm$	0.16	)	$\cdot 10^{-1}$	&	(	2.60	$\pm$	0.10	)	$\cdot 10^{-1}$	&		below detection threshold						\\
\hline
\hline																
\end{tabular}
\end{table*}
%=========================================
%==========================================
\begin{table*}
\centering
\caption{Measured angular distributions for the n$_0$, n$_1$, n$_2$, and n$_3$ transitions at $E_\alpha$ = 4.95 MeV}\label{table5}
\begin{tabular}[!h]{ccccc}
\hline \hline
$\theta_{CM}$[degrees]	&	n$_0$						&	n$_1$						&	n$_2$						&	n$_3$						\\
18.5	&	(	1.54	$\pm$	0.14	)	$\cdot 10^{-1}$	&		1.00	$\pm$	0.03			&	(	5.3	$\pm$	0.2	)	$\cdot 10^{-1}$	&	(	1.81	$\pm$	0.14	)	$\cdot 10^{-1}$	\\
27	&	(	1.21	$\pm$	0.21	)	$\cdot 10^{-1}$	&		0.91	$\pm$	0.03			&	(	5.42	$\pm$	0.17	)	$\cdot 10^{-1}$	&	(	1.87	$\pm$	0.12	)	$\cdot 10^{-1}$	\\
35.4	&	(	1.11	$\pm$	0.17	)	$\cdot 10^{-1}$	&		0.90	$\pm$	0.03			&	(	5.3	$\pm$	0.2	)	$\cdot 10^{-1}$	&	(	1.73	$\pm$	0.11	)	$\cdot 10^{-1}$	\\
43.8	&	(	0.99	$\pm$	0.12	)	$\cdot 10^{-1}$	&		0.80	$\pm$	0.03			&	(	5.35	$\pm$	0.17	)	$\cdot 10^{-1}$	&	(	1.75	$\pm$	0.13	)	$\cdot 10^{-1}$	\\
57.8	&	(	0.69	$\pm$	0.08	)	$\cdot 10^{-1}$	&		0.66	$\pm$	0.02			&	(	4.71	$\pm$	0.14	)	$\cdot 10^{-1}$	&	(	1.36	$\pm$	0.14	)	$\cdot 10^{-1}$	\\
73.2	&	(	0.43	$\pm$	0.09	)	$\cdot 10^{-1}$	&		0.48	$\pm$	0.02			&	(	3.90	$\pm$	0.11	)	$\cdot 10^{-1}$	&	(	1.24	$\pm$	0.16	)	$\cdot 10^{-1}$	\\
85.4	&	(	0.49	$\pm$	0.09	)	$\cdot 10^{-1}$	&		0.47	$\pm$	0.02			&	(	3.57	$\pm$	0.10	)	$\cdot 10^{-1}$	&	(	0.97	$\pm$	0.07	)	$\cdot 10^{-1}$	\\
93.3	&	(	0.70	$\pm$	0.09	)	$\cdot 10^{-1}$	&		0.50	$\pm$	0.02			&	(	3.59	$\pm$	0.11	)	$\cdot 10^{-1}$	&	(	0.81	$\pm$	0.08	)	$\cdot 10^{-1}$	\\
101.4	&	(	0.74	$\pm$	0.09	)	$\cdot 10^{-1}$	&		0.513	$\pm$	0.014			&	(	3.38	$\pm$	0.12	)	$\cdot 10^{-1}$	&	(	0.74	$\pm$	0.08	)	$\cdot 10^{-1}$	\\
109.3	&	(	0.75	$\pm$	0.09	)	$\cdot 10^{-1}$	&		0.53	$\pm$	0.02			&	(	3.45	$\pm$	0.19	)	$\cdot 10^{-1}$	&	(	0.71	$\pm$	0.08	)	$\cdot 10^{-1}$	\\
\hline
\hline																
\end{tabular}
\end{table*}
%=========================================
 
We  fit the data by using Legendre polynomials from P$_0$ to P$_2$ (blue dashed line in the figures) and from  P$_0$ to P$_4$ (red line).
%===========================
%===========================
%%===========================
%\begin{figure}
%\centerline{%
%\includegraphics[width=\columnwidth]{Fig8.eps}}
% \caption{The angular distribution of n$_1$ at  the $\alpha$-beam energy of 3.95 MeV are reported in black. The fit using from P$_0$ to P$_2$ Legendre polynomials is in blue dashed line, while the fit made including the Legendre polynomials until P$_4$ is reported in red line.} \label{fig:crosssection4_n1}
%\end{figure}
%%===========================
%===========================
%===========================
%===========================
%===========================
%===========================
%===========================
The first fit has been performed in the same way as discussed in the paper from Van Der Zwan and Geiger \cite{VanderZwan1981}. 
Both fits satisfactorily reproduce  the experimental data except in some cases  (see \ref{fig:crosssection3_n0} and \ref{fig:crosssection4_5}), where the best fit is obtained using higher degrees of Legendre polynomials. 

We note that the n$_1$ angular distribution reported in Fig. \ref{fig:crosssection4_5} is relative to a beam energy of 4.30$\pm$0.03 MeV.
This energy is close to the 4.295 MeV resonance corresponding to the population of the 14.829 MeV unbound level in the intermediate $^{29}$Si compound nucleus as reported by Van Der Zwan and Geiger \cite{VanderZwan1981}.
On the other hand, the n$_0$ angular distribution reported in Fig. \ref{fig:crosssection3_n0} is relative to a beam energy of 3.05$\pm$0.01 MeV that is not associated to any previously observed resonance.
Many similar resonances can be explored in the beam energy region from 3.5 to 5 MeV. 
This issue should be addressed in a future experimental campaign aimed at verifying the role of the resonances in neutron angular distributions.

%As a matter of fact, in fig. \ref{fig:crosssection4_5} the angular distribution for n$_1$ is measured with a  beam energy of 4.30$\pm$0.03 MeV. This energy is close to the 4.295 keV resonance  corresponding to the population of the 14.829 MeV unbound level in the intermediate $^{29}$Si  compound nucleus  as reported by Van Der Zwan and Geiger \cite{VanderZwan1981}.
%It has to be noted that, there are a lot of resonances in the beam energy region from 3.5 MeV to 5 MeV. Unfortunately we were not able to explore all of them due to the reduced beam current and the beam time limitations, but these, together with cross section measurements, are the aims of the future works on the \MgAl reaction.
%
%Another example where the assumptions in \cite{VanderZwan1981} do not seems to  reproduce the experimental data is  shown in fig. \ref{fig:crosssection3_n0}. This data was acquired along several days with a beam energy of 3.05$\pm$0.01 MeV.
%This energy is not associated to an already observed  resonance, but, as in the case of fig. \ref{fig:crosssection4_5}, the angular distribution could only be fitted with the red line.

In addition, looking at the experimental data it seems that in some cases the angular distribution does not  rise again at backward angles (as for the P$_0$ to P$_2$ solution), but begins to flatten above 100$^\circ$ (see fig. \ref{fig:crosssection5_n0} and fig. \ref{fig:crosssection5_n1}). A similar behaviour was observed in a study of $^6$Li($^3$He,n)$^8$B \cite{BETABEAM}. This fact leads us to speculate that the choice of using only P$_0$ to P$_2$ to parametrise the experimental data is not entirely correct, since it could overestimate the integrated total cross section.

%===========================
%===========================
\section{Conclusions}\label{sec:conclusions}
%===========================
%===========================  

The measurement of the \MgAl angular distributions at 5 different energies in the range from 3  to 5 MeV was carried out with unprecedented granularity  from \ang{17.5} to \ang{106}. 
As far as we know this is the first time that the \MgAl angular distributions   are reported considering that in previous works only the total cross section was shown.
Differences between the fitting assumptions used in \cite{VanderZwan1981} and  our experimental data were observed.

It was shown that extending the  angular coverage would be helpful to better understand the behaviour of the angular distributions at backwards angles and to improve the precision on the integrated total cross section.
It has to be noted that also in inclusive measurements where thermalised neutrons are measured \cite{Anderson1983,Wieland,Falahat2010}, the detection efficiency  could be affected by the branching probability and the angular distribution of emitted neutrons. 

Therefore, the experimental  angular distributions presented in this work imply the need of new efforts in order to obtain a more reliable and precise cross section for the \MgAl reaction. This would lead to a better understanding of the origin of the \Al in our Galaxy.
Future experimental campaign is under discussion to overcome the limitations of the present work.

%===========================
%===========================
\section*{Acknowledgments}
%===========================
%===========================
The authors are indebted to  Augusto Lombardi, Luca Maran and Lorenzo Pranovi (University of Padua) for the help in setup installation and for running the CN accelerator. We thank Massimo Loriggiola (LNL) for the target production and the INFN-PD and LNL mechanical workshops   for technical support (in particular Alessandro Minarello). A.~C. acknowledges financial support by Fondazione Cassa di Risparmio di Padova e Rovigo.

%\appendix
%\section{Experimental Angular Distribution}
%
%All values for experimental angular distributions are reported in this appendix.

\end{document}